\documentstyle [11pt]{article}
\begin{document}
\hyphenation{to-po-lo-gi-cal}
\newcommand{\af}{\mbox{$\alpha$}}
\newcommand{\be}{\mbox{$\beta$}}
\newcommand{\la}{\mbox{$\lambda$}}
\newcommand{\gh}{\mbox{$\gamma$}}
\newcommand{\de}{\mbox{$\frac{1}{2}$}}
\newcommand{\th}{\mbox{$\frac{1}{3}$}}
\newcommand{\qa}{\mbox{$\frac{1}{4}$}}
\newcommand{\sx}{\mbox{$\frac{1}{6}$}}
\newcommand{\vg}{\mbox{$\frac{1}{24}$}}

\newcommand{\npb}{{\it Nucl. Phys.}}
\newcommand{\prl}{{\it Phys. Rev. Lett.}}
\newcommand{\cmp}{{\it Commun. Math. Phys.}}
\newcommand{\jpamg}{{\it J. Phys. A: Math. Gen.}}
\newcommand{\lmp}{{\it Lett. Math. Phys.}}
\newcommand{\ptp}{{\it Prog. Theor. Phys.}}

\title{Bethe Ansatz for Lattice Analogues of $N=2$ Superconformal Theories}
\author{\vspace{1cm} Z. Maassarani\\ \small Physics
Department\\ \small University of
Southern California\\ \small Los Angeles, CA 90089-0484}
\date{ }
\maketitle
\vspace{2cm}
\begin{abstract}
\begin{normalsize}
The critical Boltzmann weights for lattice analogues of the $N=2$
superconformal coset models $\frac{G_1 \times SO(dim(G/H))}{H}$ were
given in \cite{nick}. In this paper Bethe
Ansatz methods are employed to calculate
the spectrum of the transfer matrix obtained from these Boltzmann weights.
{}From this the central charge and conformal weights are obtained by
calculating
finite-size corrections to the free energy per site. The results agree
with  those obtained from the superconformal model.
\end{normalsize}
\end{abstract}
\vspace{5cm}
\hspace{11mm}USC-93/019 \hfill \\
\vspace{8mm}
\hspace{10mm}June 1993 \hfill \\
\thispagestyle{empty}
\newpage
\setcounter{page}{1}

\section{Introduction}

The lattice analogues of $N=2$ superconformal models were constructed in
\cite{nick}. Specifically, Coulomb gas techniques were used to obtain
the critical Boltzmann weights corresponding to the models
\begin{equation}
\frac{G_1 \times SO_1(dim_{R}(G/H))}{H_{g-h+1}}
\label{coset}
\end{equation}
where $G$ is simply laced and of level one, $G/H$ is a hermitian
symmetric space and $g$ and $h$ are the dual Coxeter numbers of $G$ and $H$.
The central charge is $c=\frac{3 d}{g+1}$, where $d$ is the complex
dimension of $G/H$.
The Boltzmann weights were explicitly constructed for
the grassmannian model with $G=SU(m+n)$ and $H=SU(m)\times SU(n) \times
U(1)$. While the arguments of \cite{nick} were fairly compelling, it is
necessary to perform some fundamental checks on the results. More precisely,
the lattice model, at the critical temperature, should exhibit conformal
invariance at large distances. It is therefore the purpose of this paper to
determine the central charge and conformal weights from the Bethe Ansatz
solution to the lattice  model. The results agree with the coset model
predictions.

\vspace{5mm}

The starting point of \cite{nick} is to  consider the lattice analogues
of the  ${\cal G}_{k,1} = \frac{G_k \times G_1}{G_{k+1}}$ models.
These conformal models only appear at the critical point of
the lattice model \cite{bunch,pasquier,francesco}. In
the vertex description of the lattice model
one associates to each edge of the $45^{\circ}$ lattice a copy
of the fundamental representation, $V$, of $G$. The Hilbert space of
a vertical slice of $2 L$ edges is then $V^{\otimes 2L}$.
The continuum limit can be argued to  correspond to a model consisting
of $r$ free bosons, where $r$ is the rank of $G$.
The transfer matrix $\tau (u,q)$,
which depends on a spectral parameter
$u$ and a deformation parameter $q$, gives
the `time' evolution of this slice from left to right. It is given by
\begin{equation}
\tau (u,q) = [\prod_{p=1}^{L} X_{2p-1}(u,q)] [\prod_{p=1}^{L-1} X_{2p}(u,q)]
\label{transf}
\end{equation}
where the matrix $X_p(u,q) = \frac{1}{2i} \check{R}(u,q)$ acts on the tensor
product of the $p^{\rm th}$ and $(p+1)^{\rm th}$ copies of
$V$. The matrix $\check{R}$
is the `$\check{R}$-matrix' of $U_q (G)$. For $G=SU(n)$, $V$ is
$n$-dimensional and one has \cite{jimbo}:
\begin{eqnarray}
\check{R}(u,q)=& (xq-x^{-1}q^{-1})\sum_{k=1}^{n}
E_{kk} \otimes
E_{kk} + (x-x^{-1}) \sum_{\stackrel{k \neq l}{\scriptscriptstyle k, l= 1}}^n
E_{k l}\otimes
E_{l k}   \nonumber \\
&+ (q-q^{-1})[x \sum_{k > l} + x^{-1} \sum_{k < l}]
E_{k k} \otimes E_{l l} \: ,\: 1\leq k,l \leq n \: ,
\label{rcheck}
\end{eqnarray}
where $x=e^{iu}$ and $(E_{kl})_{i,j}=
\delta_{k, i} \, \delta_{l , j} \,$. The matrix $\check{R}$
satisfies the Yang-Baxter equation
\begin{equation}
(\check{R}(u,q)\otimes I)(I\otimes \check{R}(u+v,q))(\check{R}(v,q)\otimes I)=
(I\otimes \check{R}(v,q))(\check{R}(u+v,q)\otimes I)(I\otimes \check{R}
(u,q))\,.
\label{yb}
\end{equation}
The neareast-neighbor spin-chain Hamiltonian is given by
\begin{equation}
{\cal H}=\tau^{-1}(u)\frac{\partial \tau (u)}{\partial u}|_{u=0} \, .
\label{hamil}
\end{equation}
It contains a boundary term
\begin{equation}
{\cal H}_{bdry}=-\frac{2 i}{g} \rho_G.(h^{(1)}-h^{(2L)})
\label{hamilbdry}
\end{equation}
where $\rho_G$ is the Weyl vector of
$G$ and $h^{(j)}$ is the Cartan subalgebra generator acting on the
$j^{\rm th}$ edge of the lattice. This boundary term ensures the
commutation of ${\cal H}$ with the generators of $U_q (G)$ \cite{saleur}.
It also shifts the ground state energies and is thus the discrete counterpart
of introducing a boundary charge proportional to $\rho_G$ into a gaussian
model. One now takes $q$ such that
\begin{equation}
q^{k+g+1} = - 1 \: .
\label{unity}
\end{equation}
One can then perform the
quantum group truncation by using the modified trace
\begin{equation}
{\rm Tr}[{\cal O}]= {\rm tr}[{\cal O} \mu \otimes ... \otimes \mu ] \:,
\label{trace}
\end{equation}
where
\begin{equation}
\mu = q^{2 \rho_{G}.h}\: , \nonumber
\end{equation}
to compute the partition function and the correlation functions.
Because the transfer matrix commutes with the generators of $U_q (G)$,
it preserves this truncation.

\vspace{5mm}

In order to obtain the lattice analogue of the $N=2$ superconformal
model~(\ref{coset}) one first sets $k=0$ in ${\cal G}_{k,1}$ and
eq.~(\ref{unity}).
One then replaces the $\check{R}$-matrix by a `conjugated' one \cite{bernard},
namely
\begin{equation}
\check{R}'(x,q)\equiv [I\otimes x^{-\frac{2}{g} (\rho_G-\rho_H).h)}]\:
\check{R}(x,q) \: [x^{\frac{2}{g} (\rho_G-\rho_H).h)}\otimes I] \: .
\label{rconju}
\end{equation}
The matrix $\check{R}'$ satisfies the Yang-Baxter
equation and it commutes with $U_q (H')$ where $H'$ is the semi-simple factor
of $H$, {\it i.e. } $H=H'\times U(1)$. One should note that the foregoing
transformation is not a similarity transformation. One then only performs the
quantum group truncation with respect to $U_q (H')$. This can be implemented
by employing an $H$-modified trace to compute correlators; that is,
$\mu$ in~(\ref{trace}) is replaced by
\begin{equation}
\mu'=q^{2 \rho_H.h} \: .
\end{equation}
The conjugation of the $R$-matrix amounts to adding boundary terms
to the transfer matrix.
This is most easily seen on the Hamiltonian; ${\cal H}_{bdry}$ of
eq.~(\ref{hamilbdry}) is replaced by
\begin{equation}
{\cal H}_{bdry}'=-\frac{2 i}{g}\rho_H.(h^{(1)}-h^{(2L)})\, ,
\end{equation}
that is $\rho_G$ has been replaced by $\rho_H$. As we shall see,
this produces the shift from
$c=0$ for ${\cal G}_{0,1}$ to $c=\frac{3d}{g+1}$ for the
coset model of~(\ref{coset}).

\section{Central charge and conformal weights from the Bethe Ansatz}

The transfer matrix written in eq.~(\ref{transf})
corresponds to free boundary conditions.
This choice of boundary conditions is essential to ensure the commutation
of ${\cal \tau}$ with the generators of the quantum
group $U_q(G)$ \cite{saleur}.
However two such transfer matrices with two different spectral parameters
do not commute.
As the commutation of the transfer matrices is an essential ingredient
of the
method of the algebraic Bethe Ansatz one would like to consider periodic
boundary conditions. It was shown in \cite{saleur} how
the quantum group structure
can be exihibited for the $SU(2)$ spin-chain with twisted periodic boundary
conditions after properly choosing the twist parameters and the magnetization
of the states acted upon. I shall assume that this type of investigation
generalizes to other groups.
The use of a $90^{\circ}$ lattice with a row-to-row
transfer matrix is another ingredient of the Bethe Ansatz.
The transfer matrices differ from those corresponding to the $45^{\circ}$
lattices.
I shall consider here the continuum limit of the chains
with twisted periodic boundary conditions with a twist
matrix $\mu'$.

\vspace{7mm}

One more remark is in order before proceeding to the calculation. The
`conjugation' operation appearing in eq.~(\ref{rconju})
can be interpreted as a `gauge
transformation' \cite{sogo,mezincescu}. The transfer matrix in
eq.~(\ref{transf}) is not invariant under gauge transformations as
was seen above.
The transfer matrix constructed in \cite{mezincescu} and the
induced  Hamiltonian are for free boundary conditions and certain
surface terms. They are invariant
under gauge transformations. Hence the foregoing transfer matrix is
not equivalent to that of ref. \cite{mezincescu}. For both types
of transfer matrices the corresponding Bethe Ansatz equations
have not been written for
$SU(n)$ with $n>2$. This precludes for the moment a direct approach.

\vspace{5mm}

I shall write the Bethe Ansatz equations for  $G=SU(n)$
but the method works for other
simply laced groups and the final results are still valid with
the appropriate modifications.
The row-to-row transfer matrix with twisted periodic boundary
conditions is constructed out of a diagonal twist matrix $M$ and operator
matrices ${\cal L}_{ai}$ where
\begin{equation}
{\cal L}_{ai}=(xq-x^{-1}q^{-1})^{-1}{\cal P}_{ai}\check{R}_{ai}
\end{equation}
and $\cal{P}$ is the permutation operator on $V \otimes V$. Recall from
equation~(\ref{rcheck}) that
$\check{R}$ and hence ${\cal P}\check{R}$ act on $V \otimes V$. The index $a$
correspond to a {\it single} copy of $V$, the `auxiliary' space, and $i$ to
one copy of $V$ at each site, a `quantum' space, of the
$L$-site chain. Therefore
the operator ${\cal L}_{ai}$ acts non-trivially at the site $i$. The transfer
matrix is given by
\begin{equation}
\tau (u) = tr_a (M_a T(u))
\end{equation}
where
\begin{equation}
T(u)= {\cal L}_{a1} {\cal L}_{a2}... {\cal L}_{aL}
\end{equation}
is the `monodromy' matrix. The trace is taken on the auxiliary space.
Untwisted periodic boundary conditions correspond to a matrix $M$
proportional to the identity on $V$. From the Yang-Baxter equation~(\ref{yb})
one can derive
\begin{equation}
\check{R}(u-v,q) \: T(u) \otimes T(v) = T(v) \otimes T(u) \, \check{R}(u-v,q)
\label{rtt}
\end{equation}
where $\check{R}$ and the tensor product correspond to two copies of an
auxiliary space (see ref. \cite{rev} for a review).
A multiplication with respect to quantum indices, at each
site, is implicit. For any diagonal $M$ one has $[M\otimes M, \check{R}]=0$.
One can then easily show that two transfer matrices with different spectral
parameters commute. One can then construct a common set of
eigenvectors for these transfer matrices using the nested algebraic Bethe
Ansatz method \cite{babelon}. The
Ansatz for the eigenvector consists of a linear combination of vectors
obtained by  applying
certain `creation' operators  obtained from the monodromy matrix $T$ taken at
yet undetermined spectral parameters $u_i$ on an `initial'
eigenvector of $\tau$. The algebraic relations given in~(\ref{rtt}) are used
to obtain certain conditions on the parameters $u_i$ and on the coefficients
of the linear combination; a similar Ansatz is then made $r-2$ times
(the nesting). The complete set of conditions on the $u_i$'s constitute
the Bethe Ansatz equations.
I have modified the calculations accordingly to take into
account the twist matrix $M$.
In what follows $M$ is equal to $\mu'$ and is diagonal with elements $m_1,...,
m_n$.
Taking the logarithm of the nested Bethe Ansatz equations for
twisted periodic boundary conditions gives:
\begin{eqnarray}
\label{eqs}
 &&\sum_{j=1}^{p_{k+1}}\phi (\la_i^{(k)}-\la_j^{(k+1)},\frac{\gh}{2})
-\sum_{j=1}^{p_{k}}
\phi(\la_i^{(k)}-\la_j^{(k)},\gh)
+\sum_{j=1}^{p_{k-1}} \phi (\la_i^{(k)}-\la_j^{(k-1)},
\frac{\gh}{2}) \nonumber \\
=&& i \log (\frac{m_k}{m_{k+1}})+2\pi I_i^{(k)}\:\: ,
\: \: \: \:   1 \leq k\leq n-1 \, ,
\end{eqnarray}
where
\begin{eqnarray*}
0 =p_n \leq p_{n-1}\leq ...\leq p_0=L \: , \:
\la_j^{(k)} = - i (u_j^{(k)} - k \frac{\gh}{2}) \: ,
\: \la_i^{(0)}=0 \: , \: q=e^{i\gh}
\end{eqnarray*}
and $\phi (z,\af)$ is defined by
\begin{equation}
\phi (z,\af)= i\log \left( \frac{\sinh(z+i\af)}{\sinh(z-i\af)} \right)
\end{equation}
with $\phi (0,\af)=\pi$.
One can show that the integers $I_i^{(k)}$ are bounded; one then
solves for the roots of equations~(\ref{eqs}) for each set of integers
within the bounds.
For $0 < \gh < \frac{\pi}{2}$ one has a critical regime whose
continuum limit is gapless and corresponds to a conformal field theory.
The eigenvalues are given by:
\begin{equation}
\Lambda_M (u) = \prod_{i=1}^{p_0} \frac{1}{a(u-u_i^{(0)})}\sum_{j=1}^n m_j
\prod_{l=1}^{p_{j-1}} a(u-u_l^{(j-1)}) \prod_{m=1}^{p_j} a(u_m^{(j)}-u)
\label{lm}
\end{equation}
where $a(u)=\frac{\sin (\gh -u)}{\sin (u)}$.
The equations~(\ref{eqs}) encode the root
structure of the $A_n$ Dynkin diagram (see ref. \cite{rev} for instance).
One  has similar equations for the other Lie algebras;
they involve the simple roots
and the highest weight of the representation considered \cite{resh}.
For large $L$, fixed ratios $p_i/L$ ($i=1,..,n-1$) and $0<u<\gh/2$, one can
show that the leading term in $\Lambda_M$ is given
by the first term in~(\ref{lm}) for which $j=1$; the remaining
terms give exponentially small corrections which do not affect the finite-size
expansion in $1/L$.

The central charge and conformal weights can then be extracted following a
procedure similar to that of refs. \cite{karow,vega}.
The calculations were modified
to take into account the twist matrix contribution. One takes large values
of $L$ and then calculates finite-size corrections in $1/L$ for
the free energy per site
$f_L (u)=-\frac{1}{L} \log \Lambda_M (u)$. As $L$ becomes large the set
of solutions of the Bethe Ansatz equations take a specific distribution
in the complex plane. The exact distribution is not known. However a
conjecture supported by numerical computations is usually made for the form
of the roots of eqs.~(\ref{eqs}); it is called the `string hypothesis'.
One then postulates the existence of some  densities for the root
distribution and replaces sums by integrals in the continuum limit.
These densities satisfy a set of equations obtained from eqs.~(\ref{eqs}).
One then calculates the corrections up to $1/L^2$ for
$f_L - f_{\infty}$. The form of the  leading terms of the
$\frac{1}{L}$-expansion were predicted in \cite{cardy} by conformally mapping
the complex plane onto a strip of width $L$. The central charge and conformal
weights appear in this expansion. Upon comparing the two expansions I obtain:

\begin{eqnarray}
&c=r-\frac{12}{\pi (\pi-\gh)} \vec{t} A^{-1} \vec{t} \, , \\
\label{delta1}
&\Delta=\frac{1}{2(1-\gh/\pi)}(\vec{h}^+ -\frac{\gh}{2\pi}\vec{S}+\frac{1}{\pi}
\vec{t})A^{-1}(\vec{h}^+ -\frac{\gh}{2\pi}\vec{S}+\frac{1}{\pi}
\vec{t}) + \frac{c-r}{24} \, , \\
&\overline{\Delta}=\frac{1}{2(1-\gh/\pi)}(\vec{h}^- -\frac{\gh}{2\pi}\vec{S}
-\frac{1}{\pi}\vec{t})A^{-1}(\vec{h}^- -\frac{\gh}{2\pi}\vec{S}-\frac{1}{\pi}
\vec{t}) + \frac{c-r}{24} \, ,
\label{delta2}
\end{eqnarray}
where $t_j=\frac{1}{2i} \log(\frac{m_j}{m_{j+1}})$, $A$ is the Cartan matrix
of G and $r$ its rank, $\vec{h}^{\pm}$ are $r$-dimensional
vectors of integers labeling excitations (holes {\it and} complex strings)
and $ \de S_j = \frac{p_{j-1}+p_{j+1}}
{2} - p_j $ is the `spin' at the Ansatz-level $j$. The vectors
$\vec{h}^{\pm}$ and $\vec{S}$ are not independent, namely one must have
$\vec{h}^+ + \vec{h}^-  = \vec{S}$. The number of sites $L$ is taken to be
a multiple of $m+n$. This is a natural requirement if one considers the
ground state of the statistical model, for which $\vec{S}=\vec{h}^{\pm}=
\vec{0}$. With $p_0 = L$ and $p_L=0$ this implies $p_i=L-\frac{L}{g}i$ which
should be integers. Therefore $L$ should be a multiple of $m+n$. Intuitively,
the ground state should be made up of a  linear combination of vectors with
equal number of $SU(m+n)$ `spins', or vectors in the canonical basis of
${\bf R}^{m+n}$, in each directions; for $SU(2)$ an equal
number of spins up and spins down.
The Bethe Ansatz equations for various twisted periodic boundary
conditions for $SU(2)$ were analyzed in \cite{karow}. There, an even number
of sites was considered implying integer spin.

One sees that the central charge has been decreased from its untwisted value
of $r$.  These formulae are valid for any simply laced Lie group.

\vspace{5mm}

Consider now the lattice grassmannian models where $G=SU(m+n)$
and $H=SU(m)\times SU(n)\times
U(1)$, with $m+n \geq 3$. The twist matrix is equal to
\begin{equation}
M=q^{2\rho_H.h}= \:{\rm diag} \: (q^{m-1},q^{m-3},..,
q^{-m+1},q^{n-1},q^{n-3},..,q^{-n+1}) \: .
\label{m}
\end{equation}
With the choices made for the roots of $G$ and $H$ one has $\rho_G -\rho_H =
\frac{m+n}{2} \la_m$ where $\la_m $ is the $m^{\rm th}$ fundamental
weight of $G$.
Define $x_i\equiv \frac{1}{i\gh}\log m_i$, then
$t_i=\frac{\gh}{2}(x_i-x_{i+1})$  and one can rewrite the central charge as
\begin{eqnarray}
c & = & m+n-1-\frac{3\gh^2}{\pi(\pi-\gh)(m+n)}
\sum_{1\leq i,j\leq m+n}(x_i-x_j)^2   \nonumber \\
  & = & m+n-1-\frac{3\gh^2}{\pi(\pi-\gh)} \sum_{1\leq i \leq m+n} x_i^2
\label{cen}
\end{eqnarray}
where the last equality follows from $\sum_{i=1}^{n+m} x_i = 0$.
Since $\gh=\frac{\pi}{g+1}=\frac{\pi}{m+n+1}$, $x_i=m+1-2i$ for
$i=1$ to $m$ and $x_i=2m+n+1-2i$ for $i=m+1$ to $m+n$ one obtains
$c=\frac{3mn}{m+n+1}$. Therefore one recovers the central charge for
the corresponding coset model.

\section{The continuum model}
The conformal weights for the $N=2$ superconformal coset model of
equation~(\ref{coset})
can be read off the modified $A$-type modular
invariant gaussian partition function
\cite{nick} :
\begin{eqnarray}
Z = \frac{1}{2 \mid W(H) \mid \mid Z(G)\mid}
\frac{1}{\mid \eta (\tau)\mid^{2 r}}
\sum_{w\in W(G)}\sum_{v\in \frac{\frac{1}{\beta} M(G)^*}{\Gamma}} \sum_{u\in
\frac{\beta M(G)^*}{\Gamma}} \sum_{v_1 ,v_2 \in \Gamma} \nonumber \\
\sum_{\stackrel{\xi =0,1}
{\scriptscriptstyle \eta = 0,1}} \epsilon (w) q^
{\de (v_L +\eta s)^2} \overline{q}^{\de (v_R +
\eta s)^2} e^{-4\pi i s.(\zeta_L v_L-\zeta_R v_R)} e^{-2\pi i\xi s.(v_L -v_R)}
\: ,
\end{eqnarray}
where
\begin{eqnarray}
\Gamma = \sqrt{g(g+1)} M(G)  \\
v_L =v+u+v_1 \; , \; v_R =w(v)+u+v_2
\end{eqnarray}
and
\begin{equation}
s = \frac{1}{\sqrt{g(g+1)}}(\rho_G -\rho_H) \: .
\end{equation}
The lattices appearing in the sums are the root ($M(G)$) and weight ($M(G)^*$)
lattices of $G$.
The foregoing partition function represents ${\rm Tr}[q^{L_0 -c/24}
\overline{q}^{\overline{L}_0 - c/24} e^{-2\pi i (\zeta_L J_0 -\zeta_R
\overline{J}_0)}]$ taken over the entire Hilbert space. The sum over
$\eta =0 \: {\rm and} \:1$, correspond
to the sum over the the Neveu-Schwarz and
Ramond sectors, respectively.
The conformal weights  of  the primary fields are
therefore given by:
\begin{eqnarray}
\Delta = \de (v_L +\eta s)^2 +\frac{c-r}{24} \; \; ,\; \;
\overline{\Delta} = \de (v_R +\eta s)^2 + \frac{c-r}{24} \; .
\label{del}
\end{eqnarray}
One can write
$v_L$ and $v_R$ as follows:
\begin{equation}
v_L = \sum_{i=1,r}(\frac{1}{\beta} v_i \lambda_i +\beta u_i \lambda_i +
\frac{g}{\beta} r^{(1)}_i \af_i) \; ,\; v_R = \sum_{i=1,r}(\frac{1}
{\beta} v_i w(\lambda_i) +\beta u_i \lambda_i  +\frac{g}{\beta}
r^{(2)}_i \af_i) \; ,
\label{vs}
\end{equation}
where $\vec{v} \, ,\, \vec{u} \, , \, \vec{r}_1 \,$
and $\vec{r}_2$ belong to ${\bf Z}^r$,
the $\af_i$'s are the simple roots
of $G$ ($\af_i^2=2$) and the $\lambda_i$'s are the fundamental weights of $G$.
While equations~(\ref{delta1}), (\ref{delta2}) and equations~(\ref{del})
are of an identical form, it must be remembered  that the various component
parts are constrained integer vectors.
There does not seem to be a general way of identifying the integer vectors
$\vec{v} \, ,\, \vec{u} \, , \, \vec{r}_1 \,$  and $\vec{r}_2$ with the
vectors $\vec{S} \, ,\, \vec{h}^{\pm}$ and $\vec{t}$ in the Bethe Ansatz.
One can  consider particular scalar fields
($\Delta=\overline{\Delta}$) for which $\vec{S}=0=\vec{h}^+ +\vec{h}^-$.
The vectors $\vec{h}^{\pm}$ can be negative because they label holes
{\it and} string excitations.
Using $\lambda_i . \af_j = \delta_{i,j}\,$, $\lambda_i = A^{-1}_{i j}\af_j$
and the symmetry of the Cartan matrix for simply laced groups it is easy to
see that the weights~(\ref{delta1}) and~(\ref{delta2}) can be put in the
form~(\ref{del}) provided that:
\begin{eqnarray}
v_i=h^+_i + \frac{x_i-x_{i+1}-(m+n)\delta_{i,m} \eta}{2} - k_i g \: ,
\: u_i= \frac{-x_i+x_{i+1}+(m+n)\delta_{i,m} \eta}{2}
+k_i (g+1) \nonumber \\
r^{(1)}_i=r^{(2)}_i=0 \: ,\: k_i \in {\bf Z} \: ,
\label{sol}
\end{eqnarray}
where $\eta =0$ (the NS sector) if $m+n$ is even and $\eta=1$ (the Ramond
sector) if $m+n$ is odd.
Indeed, for even $m+n$
the twist integers $(x_i -x_{i+1})$ are even as can be seen from the matrix
$M$ in eq.~(\ref{m}). For
odd $m+n$ only $x_m -x_{m+1}$ is odd.

\vspace{5mm}

The Bethe Ansatz conformal weights~(\ref{delta1}) and~(\ref{delta2})
were found to be identical in form to those of the grassmannian coset model.
Specific subsets of the set of weights of the coset model can be obtained
from the Bethe Ansatz conformal weights. This is another confirmation
that one has a lattice analogue of the coset model.
The observation that there does not seem to be a general way
of identifying the components of the coset model conformal weights with
the parameters of the Bethe Ansatz weights is not worrying.
One has no reason to expect  the weights obtained from the Bethe Ansatz
to be organized exactly as those of the coset model. Indeed the Bethe
Ansatz was done for twisted periodic boundary conditions instead of free
boundary conditions and the analysis of the excitations does not
seem naturally related to the coset model labels.

\section{Conclusion}
The Bethe Ansatz method and finite-size techniques were used to study
the continuum limit of a lattice that should exhibit
at criticality in the continuum
limit an $N=2$ superconformal behaviour.
The central charge obtained
from the Bethe Ansatz
is identical with that of the coset model to which the lattice model
is expected to flow in the continuum limit. The conformal
weights obtained for the  specific  set of excitations considered,
correspond to a subset of the full set of weights of the $N=2$
superconformal theory~(\ref{coset}).
This confirms the identification of the finite lattices at the critical
temperature with the $N=2$ superconformal models to which they are expected
to flow in the continuum limit.
Twisted periodic boundary conditions
were used assuming that an analysis similar to that of ref. \cite{saleur}
for $SU(2)$ generalizes to $SU(n)$ with $n>2$. Therefore one
expects an overlap between the set of states of the twisted periodic transfer
matrix and those of the transfer matrix~(\ref{transf}) constructed
from $\check{R}'$. Furthermore Bethe Ansatz equations have not yet
been derived for free boundary conditions and surface terms for
$SU(n)$ with $n>2$. Such an Ansatz would provide the full set of conformal
weights corresponding to the coset model. It would then  provide another
valuable verification that one has a lattice analogue of an $N=2$
superconformal coset model.
A further and delicate step in this
identification consists of studying the operator content, including
the operator multiplicities, of the statistical
model.

\vspace{1cm}
\hspace{-6mm}\Large\bf Acknowledgement \\
\normalsize
\\
I would like to thank N. Warner for encouragement and numerous discussions
and H. Saleur for discussions.


\begin{thebibliography}{30}
\bibitem{nick} Z. Maassarani, D. Nemeschansky and N. P. Warner,
{\it Nucl. Phys.} {\bf B393}, (1993) 523.
\bibitem{bunch} A.A. Belavin, \npb~{\bf B188}, (1981) 189; O. Babelon,
H. J. de Vega and C. Viallet, \npb~{\bf B190}, (1981) 542; P.P. Kulish,
N.Yu. Reshetikhin and E.K. Sklyanin, \lmp~{\bf 5}, (1981) 393; M. Jimbo,
T. Miwa and M. Okada, \lmp~{\bf 14}, (1987) 123; \cmp~{\bf 116}, (1988) 507;
M. Jimbo, A. Kuniba, T. Miwa and M. Okada, \cmp~{\bf 119}, (1988) 543.
\bibitem{pasquier} V. Pasquier, \npb~{\bf B295}, (1988) 491.
\bibitem{francesco} P. Di Francesco and J.B. Zuber, \npb~{\bf B338},
(1990) 602.
\bibitem{jimbo} M. Jimbo, {\it Commun. Math. Phys.} {\bf 102}, (1986) 537.
\bibitem{bernard} D. Bernard and A. Le Clair, {\it Commun. Math. Phys.}
{\bf 142}, (1991) 91.
\bibitem{saleur} V. Pasquier and H. Saleur, {\it Nucl. Phys.} {\bf B330},
(1990) 523.
\bibitem{sogo} K. Sogo, Y. Akutsu and T. Abe, \ptp~{\bf 70}, (1983) 730.
\bibitem{mezincescu} L. Mezincescu and R. I. Nepomechie, {\it J. Phys. A:
Math. Gen.} {\bf 24}, (1991) L17.
\bibitem{babelon} O. Babelon, H. J. de Vega and C.-M. Viallet,
{\it Nucl. Phys.} {\bf B200}, (1982) 266.
\bibitem{rev} H. J. De Vega, {\it Int. J. Mod. Phys.} {\bf A10}, (1989) 2371.
\bibitem{karow} M. Karowski, \npb~{\bf B300}, (1988) 473.
\bibitem{vega} H. J. de Vega, {\it J. Phys. A: Math. Gen.} {\bf 21},
(1988) L1089; H. J. de Vega and M. Karowski, {\it Nucl. Phys.} {\bf B285},
(1987) 619; H. J. de Vega and F. Woynarovich,
{\it Nucl. Phys.} {\bf B251}, (1985) 439.
\bibitem{cardy} J. L. Cardy, {\it J. Phys. A: Math. Gen.} {\bf 19}, (1984)
L1093; I. Affleck, {\it Phys. Rev. Lett.} {\bf 56}, (1986) 746;
H. W. J. Bl\"{o}te, J. L. Cardy and M. P. Nightingale,
{\it Phys. Rev. Lett.} {\bf 56}, (1986) 742.
\bibitem{resh} N. Yu. Reshetikhin, \lmp~{\bf 14}, (1987) 235.
\end{thebibliography}
\end{document}